\documentclass[a4paper,11pt]{article}
\usepackage[english]{babel} 
\usepackage[T1]{fontenc}
\usepackage{soul}
\usepackage{float}
\usepackage{setspace}
\usepackage{appendix}
\usepackage{parcolumns}
\usepackage{stmaryrd}
\usepackage{cite}
\usepackage{times}
\usepackage[OT1]{fontenc}
\usepackage{type1cm}
\usepackage{url}
\usepackage{subfigure}
\usepackage{braket}
\usepackage{graphicx}
\usepackage{sidecap}
\usepackage{xspace}
\usepackage{tikz}
\usepackage{pgflibraryarrows}
\usepackage{pgflibrarysnakes}
\usepackage{indentfirst}
\usepackage[mathscr]{eucal}
\usepackage{geometry}
\usepackage{booktabs}
\usepackage{indentfirst}
\usepackage{bm}
\usepackage{multirow}
\usepackage{dcolumn}
\usepackage{graphicx}
\usepackage{mathrsfs}  
\usepackage{enumerate}
\usepackage[utf8]{inputenc}
\usepackage{amsmath}
\usepackage{amssymb}
\usepackage{amsfonts}
\usepackage{amsthm}
\usepackage{mathrsfs}
\usepackage{makeidx}
\usepackage{graphicx}
\usepackage{ulem}			

\DeclareMathOperator{\arccot}{arccot}

\makeatletter
\@addtoreset{equation}{section}

\makeatother

\def\df{{d_\phi}}
\newcommand{\be}{\begin{equation}}
\newcommand{\ee}{\end{equation}}
\newcommand{\bea}{\begin{eqnarray}}
\newcommand{\eea}{\end{eqnarray}}
\newcommand{\vs}[1]{\vspace{#1 mm}}

\def\cO{{\cal O}}

\begin{document}

\vs{8}
\begin{center}
{\Large\bf Renormalization Group flows\\ 
between Gaussian Fixed Points}
\vs{6}

{\large
Diego Buccio\footnote{e-mail address: dbuccio@sissa.it}\ 
Roberto Percacci\footnote{e-mail address: percacci@sissa.it}\ 
\vs{3}

{SISSA, International School for Advanced Studies, via Bonomea 265, 34136 Trieste, Italy}
{ and\\ INFN, Sezione di Trieste, Italy}
}

\vs{8}
{\bf Abstract}
\end{center}
A scalar theory can have many Gaussian (free) fixed points,
corresponding to Lagrangians of the form $\phi\,\Box^k\phi$.
We use the non-perturbative RG to 
study examples of flows between such fixed points.
We show that the anomalous dimension changes continuously
in such a way that at the endpoints the fields have
the correct dimensions of the respective free theories.
These models exhibit various pathologies,
but are nonetheless interesting as examples of theories
that are asymptotically free both in the infrared
and in the ultraviolet.
Furthermore, they illustrate the fact that a diverging
coupling can actually correspond to a free theory.

\section{Introduction}

Perturbative methods are powerful tools to study
the properties of quantum or statistical
field theories in the neighborhood of a fixed point (FP)
but they do not say much about the global properties
of the theory space. For example, one would like to
know which FP can be joined by an RG 
trajectory to another FP.
Such questions can sometimes be answered,
for example by the $c$-theorem in two dimensions
or the $a$-theorem in four.
Another possibility is to simply solve the RG equations.
This is impossible in the full theory space,
but it can be done within approximations.
For example, in the $\mathbb{Z}_2$-invariant scalar theory
in three dimensions, one can find trajectories that join
the (free) Gaussian FP in the UV to the 
Wilson-Fisher (WF) FP in the IR.

There are cases that would seem more trivial,
but that are harder to visualize.
For example, remaining in the context of $\mathbb{Z}_2$-invariant scalar theory in four dimensions, 
one can think of infinitely many
Gaussian FP's corresponding to the Lagrangians
$\phi\,\Box^k\phi$.
We will refer to them as GFP$_k$.
Most of these do not correspond to unitary theories 
in Minkowski space
but they still make sense in Euclidean space as statistical models.
Each of them can be viewed as sitting in the origin of theory space,
but then all the others are nowhere to be seen.
Which GFP we see is related to which GFP we take as the
starting point of a perturbative expansion,
and hence to the canonical dimension of the field.
For example (in four dimensions) 
GFP$_1$ is in the origin of a theory space
for a field of canonical dimension one,
GFP$_2$ in the origin of theory space
when the field is dimensionless, and so on.
In this way it would almost seem that for each choice 
of field dimension we have a different theory space,
and that these spaces are unrelated to each other.
There is some physical basis for this point of view,
because different GFP's have different numbers of
propagating degrees of freedom.
One could view the theory space
where a given GFP is in the origin as describing
the interactions of a particular set of physical degrees of freedom.
For example, whereas in Minkowski space
GFP$_1$ describes a single propagating
scalar degree of freedom, GFP$_2$ describes two.
When GFP$_2$ is infinitesimally deformed by adding
a term of the form $\phi\,\Box\phi$,
one of the two fields is massive and
the other is massless.
By integrating out the massive degree of freedom
one remains with the free massless one.
Thus there should be an RG trajectory joining GFP$_2$ in the UV
to GFP$_1$ in the IR.

There is one obvious trajectory that does this:
it consists of ``generalized free theories'' with Lagrangians
of the form
\be
\frac12 \phi\left(Z_1\Box+Z_2\Box^2\right)\phi\ .
\label{ct}
\ee
In the RG one has to parametrize the theory space with
dimensionless coordinates,
and if we choose for example the field to have canonical
dimension of mass, $Z_1$ is already dimensionless,
and the other direction is parametrized by $\tilde Z_2=Z_2 k^2$,
where $k$ is some external ``renormalization group scale'',
that in the present situation we can identify with the momentum $p$.
In these free theories, $Z_1$ and $Z_2$ do not run,
so $\tilde Z_2$ is negligible at low energy,
but dominant at high energy.
Note that choosing a different dimension for the field does not
change this conclusion.
\footnote{We show in general in Appendix \ref{app:redef} that physical properties
of the the RG flow are independent of the choice of dimension of the field.}
For example, if the field is dimensionless,
$Z_2$ is already dimensionless and the other direction
has to be parameterized by $\hat Z_1=Z_1/k^2$.
So, again, $Z_1$ is dominant at low energy
and negligible at high energy.
In both cases, this ``classical RG'' just tells us that 
the four derivative
term dominates over the two-derivative term at high energy.

We are interested in trajectories that go through
interacting regions of theory space.
In this paper we will focus mainly on 
a shift-symmetric and $\mathbb{Z}_2$-symmetric scalar field.
These symmetries restrict the Euclidean free energy,
or effective action, to have the form
\be
\Gamma[\phi]=\int d^4x\left[
\frac12 Z_1(\partial\phi)^2
+\frac12 Z_2(\Box\phi)^2
+\frac14 g((\partial\phi)^2)^2
+\ldots\right]
\label{action1}
\ee
where the ellipses stand for terms with six or more derivatives.

In perturbation theory, one would normally consider two cases: 
either $Z_1=0$ or $Z_2=0$.
These CFTs have been discussed recently in \cite{Safari:2021ocb}.
Beyond perturbation theory,
the action (\ref{action1}) and its generalizations
containing higher derivative terms,
are part of a single ``theory space'',
where all terms can be present simultaneously.
One is then interested in understanding the mutual relations
between different fixed points.
In particular, the question we shall investigate is whether
there exist nontrivial RG trajectories joining them.

The tool we shall use is the Wetterich-Morris form of the 
non-perturbative RG equation for the 1-PI effective action,
a.k.a. the effective average action (EAA)
\cite{wett1,Morris:1993qb}.
The EAA is a functional of the fields depending on an external 
scale $k$ that acts as an IR cutoff.
By making an ansatz for the EAA of the form (\ref{action1}),
the constants $Z_1$, $Z_2$ and $g$ become $k$-dependent
running couplings. Inserting the ansatz in the RG equation 
one can read off the beta functions and anomalous dimensions.

We shall calculate the RG flow based on two different choices
of field dimension, which are in turn related to different
Gaussian FP's, and show how these flows are related by a
coordinate transformation in theory space.
This yields a global picture of the flow where
both GFP's are simultaneously present.
In the neighborhood of a Gaussian fixed point,
the anomalous dimension must be small.
If, following the RG flow, we end at another FP,
we can in principle calculate the anomalous dimension of the field
at this endpoint.
Such calculations are always based on some approximations
and therefore the calculated anomalous dimension is 
generally not exact.
Remarkably, we shall see that in the case of flows between 
Gaussian FPs the result is exact, in the following sense:
the canonical dimension of the field at the UV FP plus the calculated
anomalous dimension gives exactly the canonical dimension of
the field at the IR FP.

These are the main results of the paper and occupy Section 2.
In Section 3 we discuss the trivial FP with $k=0$.
Further discussion of the results is contained in Section 4.
Some appendices contain
technical results concerning the
choice of regulator in the definition of the RG,
and the effect of non-canonical field dimensions on the RG flow.

\section{Flowing from GFP$_2$ to GFP$_1$.}

Implicit in the definition of free particle states is the choice of
a Gaussian FP.
Then, it is natural to give the field the canonical dimension that
pertains to that free theory.
When one contemplates flows interpolating between
different FP's, the choice of dimension of the field
is no longer so natural.
In this section we will discuss flows joining GFP$_1$ and GFP$_2$,
where the fields have canonical dimension one and zero respectively.
We will therefore exhibit the flow equations in both cases.
The power counting in the two cases is very different,
but the calculation of the loop contributions is essentially the same.
\footnote{This point is discussed in general in Appendix \ref{app:redef}.}
The differences arise from the choice of dimensionless coordinates
for theory space, that come natural in the neighborhood of each FP.
For each GFP$_i$ ($i=1,2$) we will therefore define a chart,
consisting of an open subset of theory space $U_i$ 
and suitable coordinate functions.
We will then discuss the transformation between the two charts
and give a global picture of the RG flow.

In order to write an explicit RG equation we have to
choose a form for the cutoff (or ``regulator'')
function $R_k$ that suppresses
the low momentum modes in the path integral.
We choose:
\be
R_k^{(24)}=Z_1(k^2-q^2)\theta(k^2-q^2)+Z_2 (k^4-q^4)\theta(k^4-q^4)\ .
\label{cut12}
\ee
The presence of the couplings $Z_1$ and $Z_2$ makes it
an ``adaptive'' cutoff
(in contrast to a ``non-adaptive'' or 
``pure'' cutoff \cite{Narain:2009qa}).
Normally only one of the two terms is considered,
but for our purposes this choice is preferable,
because it treats the two possible kinetic terms on an equal footing.
Additionally, it leads to the simplest beta functions,
among the choices we have tried.
We discuss in the appendix different choices of the cutoff.

\subsection{Dimensionful field and the chart $U_1$}
\label{sec:U1}

If in the action (\ref{action1})
we choose the term with two derivatives to define the propagator,
the field $\phi$ has canonical dimension of mass.
Then $Z_1$ is dimensionless, $Z_2$ has dimension of inverse mass squared and $g$ of inverse mass to power 4.
the power counting is that of a non-renormalizable theory.
It is natural to parametrize theory space by
\be
\tilde g=\frac{g k^4}{Z_1^2}\ ,\qquad
\tilde Z_2=\frac{Z_2 k^2}{Z_1}\ .
\label{defdimful}
\ee
The powers of $k$ make the couplings dimensionless
and the powers of $Z_1$ are such that if
these definitions are inserted in the action,
$Z_1$ can be set to one by rescaling the field.
This makes it clear that $Z_1$ is a redundant coupling.
The anomalous dimension $\eta_1=-\partial_t\log{Z_1}$
(where $t=\log(k/k_0)$) is
\be
\eta_1=
\frac{8\tilde g\left(1+2\tilde Z_2\right)
}{\tilde g+128\pi^2(1+\tilde Z_2)^2}
\label{eta1sfull24}
\ee
and the beta function of $g$ is
\bea
\beta_{\tilde g}&\equiv&\partial_t\tilde g=
(4+2\eta_1)\tilde g
+\frac{10+20\tilde Z_2-\eta_1}{64\pi^2(1+\tilde Z_2)^3}
\tilde g^2
\ .
\label{betagetafull24}
\eea
where (\ref{eta1sfull24}) has to be used.
The beta function of the dimensionful $Z_2$ vanishes,
which implies
\be
\beta_{\tilde Z_2}\equiv\partial_t\tilde Z_2
=(2+\eta_1)\tilde Z_2\ .
\label{betaZ2}
\ee

These beta functions have some nontrivial zeroes.
We see from (\ref{betaZ2}) that $\beta_{\tilde Z_2}$ can 
vanish in two ways: one is by having $\tilde Z_2=0$,
the other by $\eta_1=-2$.
Besides GFP$_1$ there are two FPs of the first type,
occurring at 
$\tilde g=128\pi^2 (2 \sqrt 6-5)$
and
$\tilde g=128\pi^2 (-2 \sqrt 6-5)$
and one FP of the second type at
$\tilde Z_2=-3/5$,
$\tilde g=-512\pi^2/5$.
The properties of these FPs are summarized
in Table \ref{tab:FPdimful}.
Some of these FPs had already been observed in 
\cite{deBrito:2021pyi,Laporte:2021kyp,Laporte:2022ziz}.

We note that the beta functions have singularities
for $\tilde g=-128\pi^2(1+\tilde Z_2)^2$ and for $\tilde Z_2=-1$.
The fixed points GFP$_1$, NGFP$_1$ are on one side of
the singularities,
while the other two are on the other side.
Thus for the purpose of studying the flows that can
start/end at GFP$_1$, the area with $\tilde Z_2<-1$
or $\tilde g<-128\pi^2(1+\tilde Z_2)^2$
is unphysical.

\begin{table}
\begin{center}
\begin{tabular}{lccccc} 
\toprule
FP & $\tilde Z_{2*}$ & $\tilde{g}_*$ & $\eta_{1*}$&$\theta_1$&$\theta_2$ \\
\midrule
 GFP$_1$& 0 & 0&0&-4&-2 \\
 NGFP$_1$ &0&-127.6&-0.90&4.40&-1.10\\
 NGFP$_2$ &0&-12505&8.90&43.60&-10.90\\
 NGFP$_3$ &-0.6&-1011&-2&-13.84&10.84\\
\bottomrule
\end{tabular}
\end{center}
\caption{Finite FPs seen with dimensionful field.}
\label{tab:FPdimful}
\end{table}

One can get a general overview picture of the flow 
in the chart $U_1$ by defining
\bea
\tilde Z_2&=&\tan u\\
\tilde g &=& 128\pi^2 (2 \sqrt 6-5)\tan v\ .
\eea
The rescaling factor has been chosen in such a way that
NGFP$_1$ is at $u=0$, $v=\pi/4$,
while the singularity of the flow is at
$u=-\pi/4$.

If we study the function $\eta_1$ in the bottom right quadrant,
we find that the condition $\eta_1=-2$ is satisfied asymptotically for
$\tilde Z_2\to\infty$ and
\be
\tilde g\sim -16\pi^2\tilde Z_2\ .
\label{asymp}
\ee
This leads us to suspect the existence of another FP
in the bottom right corner, outside the domain of this chart.
We also note the existence of a ``separatrix'':
the RG trajectory that arrives at NGFP$_1$
from the irrelevant direction,
corresponding to the eigenvector with components
\be
\left(\frac{5(11\sqrt 6-27)}{256\pi^2(505\sqrt 6-1237)}\epsilon,-\epsilon\right)
\approx (0.0143\epsilon,-\epsilon)\ .
\label{irrel}
\ee
This trajectory can be found numerically
and it has the asymptotic behavior (\ref{asymp}).
In fact all other trajectories in the fourth quadrant 
that end at GFP$_1$ have this same asymptotic behavior,
as we shall show later.
If we follow these trajectories in the sense of increasing $k$ or $t$,
those trajectories that emerge from GFP$_1$ at a steeper angle
reach this behavior sooner, while those that come out
nearly horizontally only reach this regime at higher $k$.
We shall discuss the meaning of these facts later.

\begin{figure}[h]
\begin{center}
\includegraphics[scale=0.62]{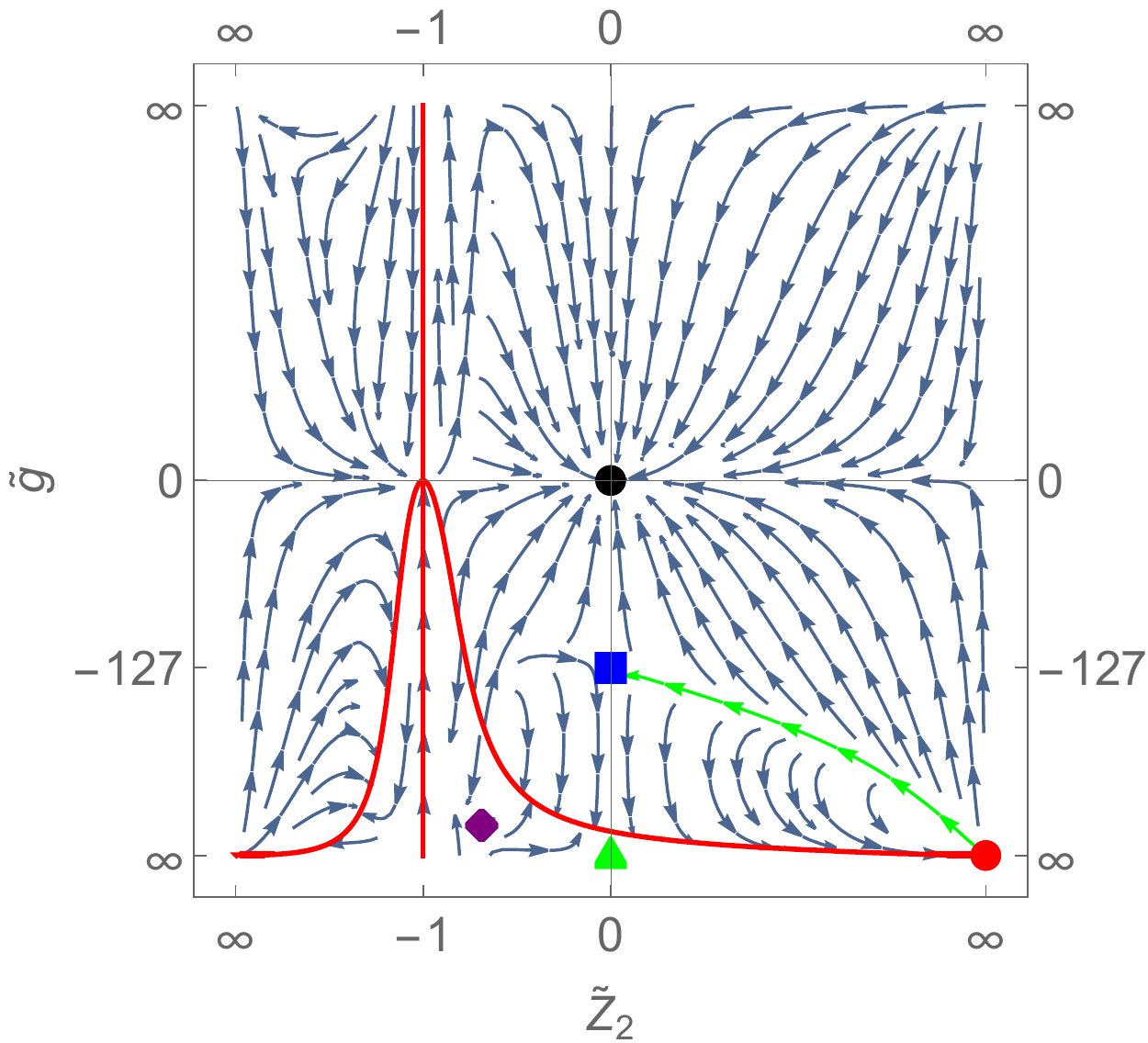}\quad
\includegraphics[scale=0.54]{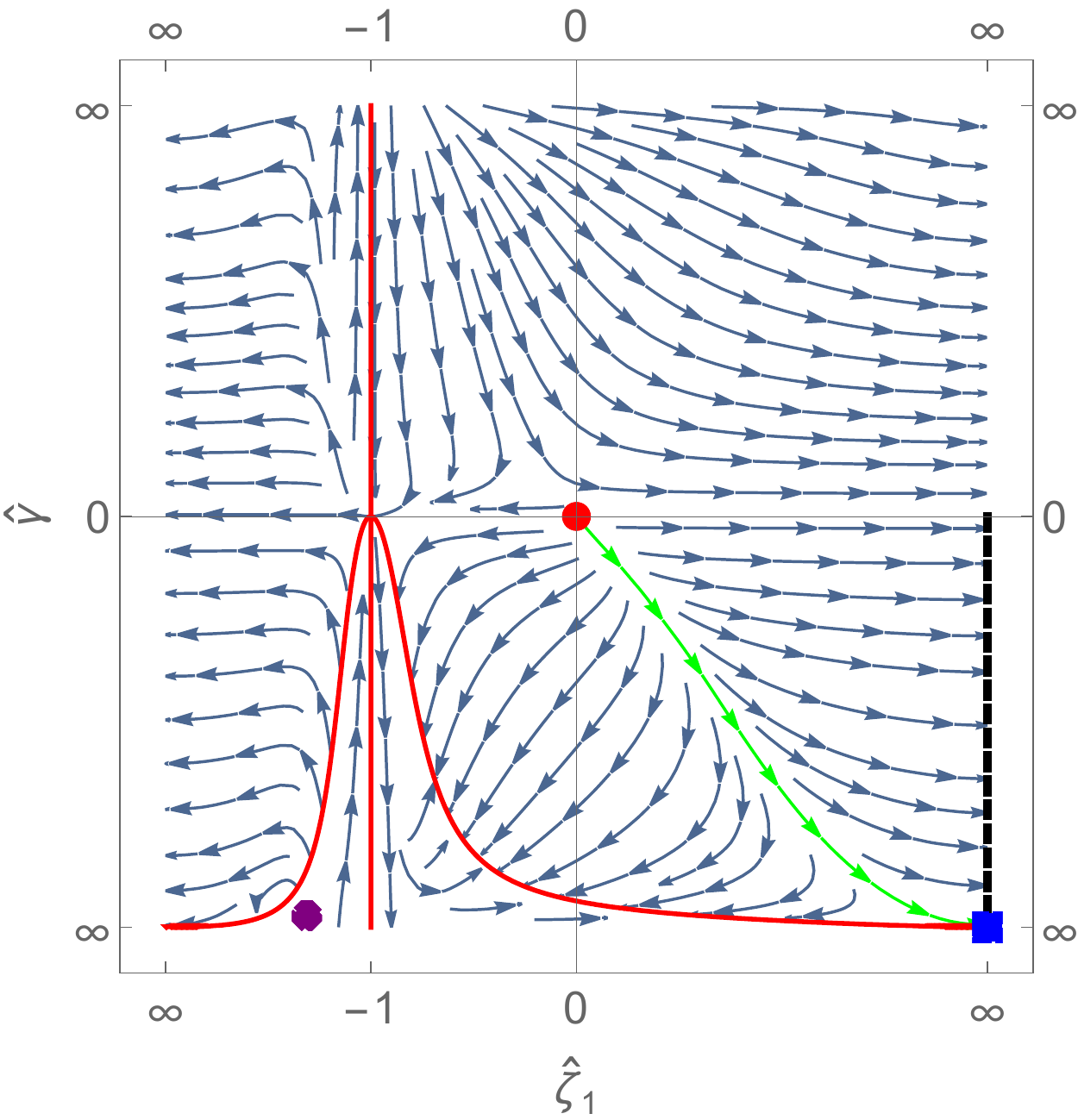}
\caption{The flow in the chart $U_1$ (left)
and in the chart $U_2$ (right).
The black dot (on the left) and the dahed black line
(on the right) mark GFP$_1$,
the red dot marks GFP$_2$,
the blue square marks NGFP$_1$.
The separatrix is the green flow line.
The continuous red lines are singularities of the flow.
}
\label{fig:flows}
\end{center}
\end{figure}

\subsection{Dimensionless field and the chart $U_2$}
\label{sec:U2}

We start again from (\ref{action1}), but we assume 
that the propagator is
given by the four derivative kinetic term,
so the field is canonically dimensionless.
In order not to confuse the couplings of this case with those
of the previous section, we shall use a different notation
for the effective action:
\be
F[\varphi]=\int d^4x\left[
\frac12 \zeta_1(\partial\varphi)^2
+\frac12 \zeta_2(\Box\varphi)^2
+\frac14 \gamma((\partial\varphi)^2)^2
+\ldots\right]
\label{action2}
\ee
It can be obtained from (\ref{action1}) by changing the
dimension of the field and redefining the couplings
\be
\phi=k\varphi\ ,\quad
Z_1=k^{-2}\zeta_1\ ,\quad
Z_2=k^{-2}\zeta_2\ ,\quad
g=k^{-4}\gamma\ .
\label{redef}
\ee
Now the wave function renormalization constant $\zeta_2$
is dimensionless, while $\zeta_1$ has dimension of mass squared,
and the coupling $\gamma$ is dimensionless.
The power counting is that of a renormalizable theory,
with $\zeta_1$ having the meaning of a mass.
The natural variables for the parametrization of theory space are
\be
\hat\zeta_1=\frac{\zeta_1}{\zeta_2 k^2}\ ,
\qquad
\hat\gamma=\frac{g}{\zeta_2^2}\ .
\label{defdimless}
\ee
Inserting these definitions in the action,
we see that $\zeta_2$ is a redundant coupling,
because it can be set to one by a field rescaling.

As in the previous section, $\partial_t\zeta_2=0$, 
\footnote{There may seem to be a contradiction between this
statement, $\partial_t Z_2=0$ and the transformation (\ref{redef}).
The point is that in the derivation of the beta functions
from the functional RG flows $\partial_t\Gamma[\phi]$
and $\partial_t F[\varphi]$ one always interprets the field
as being independent of $k$. This is discussed in greater
generality in Appendix \ref{app:redef}.}
so
\be
\eta_2\equiv-\partial_t\log Z_2=0\ .
\ee
The beta functions of the variables
$\hat\zeta_1$ and $\hat\gamma$ are
\be
\beta_{\hat\zeta_1}=-2\hat\zeta_1
-\frac{8\hat\gamma(2+\hat\zeta_1)}{\hat\gamma+128\pi^2(1+\hat\zeta_1)^2}
\label{betaZ1h}
\ee
and
\be
\beta_{\hat\gamma}
=\frac{(2+\hat\zeta_1)(\hat\gamma+640\pi^2(1+\hat\zeta_1)^2)}{32\pi^2(1+\hat\zeta_1)^3
\left(\hat\gamma+128\pi^2(1+\hat\zeta_1)^2\right)}\hat\gamma^2\ .
\ee
The fixed points of these beta functions are
listed in Table \ref{tab:FPdimless}.
We note that the nontrivial FP has the same scaling exponents
as NGFP$_3$ and has been labelled accordingly.
We shall soon understand this identification better.

By expanding the beta functions in $\hat\zeta_1$ and $\hat\gamma$,
and demanding that they point towards the origin,
we find that the only direction by which one can
tend to GFP$_2$ is $(\epsilon,-16\pi^2\epsilon)$.
Also in this case in the fourth quadrant there is a separatrix.
It distingushes curves for which $\hat\zeta_1$ tends to infinity
in the limit $k\to 0$ from those for which $\hat\zeta_1$
reaches a maximum and then turns down again.
For large $\hat\zeta_1$ the separatrix has the asymptotic form
$$
\hat\gamma=-128\pi^2(2\sqrt 6-5)\hat\zeta^2\ .
$$
This limit corresponds to GFP$_1$.

\begin{table}
\begin{center}
\begin{tabular}{lccccc} 
\toprule
 FP & $\hat\zeta_{1*}$ & $\hat\gamma_*$ & $\eta_{2*}$&$\theta_1$&$\theta_2$ \\
\midrule
 GFP$_2$& 0 & 0&0&2&0\\
 NGFP$_3$&-1.67&-2807&0&-13.84&10.84\\
\bottomrule
 \end{tabular}
\end{center}
\caption{Finite FPs seen with dimensionless field.}
\label{tab:FPdimless}
\end{table}

\subsection{Global properties of the flows}

We shall now show that the two flows described
in the previous subsections are merely coordinate
transformation of each other, outside the two
Gaussian FP's, and derive various physical properties
of the system.

\subsubsection{The coordinate transformation}

The chart $U_1$ contains GFP$_1$ but not GFP$_2$, and vice-versa.
In order to understand the flows from one Gaussian FP to the other,
we must understand that in each chart ``the other'' FP is a limiting set.
To this end, we need the coordinate transformation.
From the relations (\ref{defdimful}), (\ref{redef}) and (\ref{defdimless})
the two sets of coordinates for theory space are related by
\be
\hat\zeta_1=\frac{1}{\tilde Z_2}\ ,\qquad
\hat\gamma=\frac{\tilde g}{\tilde Z_2^2}\qquad
\mathrm{or\ conversely}\quad
\tilde g=\frac{\hat\gamma}{\hat\zeta_1^2}\ .
\label{mapinv}
\ee
From here we see that in the chart $U_2$,
taking the limit $\hat\zeta_1\to\infty$
for any fixed and finite $\hat\gamma$, 
gives $\tilde Z_2=0$ and $\tilde g=0$.
Therefore, all these limit points correspond to GFP$_1$.
Conversely in the chart $U_1$,
if we take the limit $\tilde Z_2$ and $\tilde g\to\infty$
with relation (\ref{asymp}),
we find $\hat\zeta_1=0$, $\hat\gamma=0$,
which corresponds to GFP$_2$.
While mathematically clear, these statements may sound a bit
puzzling:
from the point of view of the chart $U_2$, how can it be that
the theory becomes free in the IR even as the coupling $\hat\gamma$ remains constant?
Even more dramatically, from the point of view of the chart $U_1$, 
how can it be that the theory becomes free in the UV even as 
the coupling $\tilde g$ diverges?
We shall gain a better understanding of these statements
by studying the properties of the RG trajectories.
After the picture in dimensionless variables has been clarified,
we will discuss the picture in dimensionful variables.

\subsubsection{The mass threshold}

In the chart $U_2$, $Z_1$ is a mass squared and therefore 
there is an obvious mass threshold located where $\hat\zeta_1=1$,
with the UV located to its left and the IR to its right.
By the same token, in the chart $U_1$ the mass threshold is at
$\tilde Z_2=1$, with the IR to its left and the UV to its right.
It is somehow natural to use the chart $U_2$ for energy
scales above the threshold and the chart $U_1$ for
energy scales below it, even though the validity of both charts
extends far below this point.

\subsubsection{Perturbative vs strongly interacting trajectories}

We have shown in Section \ref{sec:U2} that all trajectories
emerge from GFP$_2$ with $\hat\gamma=-16\pi^2\hat\zeta_1$.
Applying the transformation (\ref{mapinv})
this implies that in the chart $U_1$
all the trajectories have the asymptotic behavior (\ref{asymp}),
as mentioned in Section \ref{sec:U1}.

Next note that the lines with $\tilde g=0$ (in the chart $U_1$)
and  $\hat\gamma=0$ (in the chart $U_2$)
correspond to the classical trajectory (\ref{ct}),
joining GFP$_2$ in the UV to GFP$_1$ in the IR, and
consisting entirely of free theories.
At the other extreme, the separatrix is 
in some sense the ``strongest interacting'' trajectory.
In the chart $U_1$ it consists of two segments:
first the line going from GFP$_2$ in the UV
to NGFP$_1$ in the IR,
and then the trajectory with $\tilde Z_2=0$,
joining NGFP$_1$ in the UV to GFP$_1$ in the IR.
\footnote{Strictly speaking this should be seen as
two separate trajectories, since each one takes infinite RG time,
but it can be seen as the limit of trajectories
joining directly GFP$_2$ to GFP$_1$.} 
We will be interested in the infinitely many
trajectories that are contained between these two extremes,
see Figure \ref{fig:flows}.

There are trajectories
that remain entirely in the perturbative domain,
i.e. are close to the classical trajectory.
This is not obvious when one works in a fixed chart,
because both $\tilde g$ and $\hat\gamma$ do not go to zero at
both ends of the trajectory.
Flowing out of GFP$_2$ in the chart $U_2$ they are the ones for which 
$\hat\gamma$ is small at least down to the mass threshold 
$\hat\zeta_1=1$.
Eventually, if one proceeds further towards the IR,
$\hat\gamma$ remains constant.
However, around $\hat\zeta_1=1$, one can change chart:
at that point $\tilde g=\hat\gamma$ is small
and following the flow towards the IR 
in the chart $U_1$, the coupling $\tilde g$ tends to zero.
The qualitative behavior of the trajectories is
the same also when the coupling midflows is strong.

\subsubsection{The dimension of the field is automatically adjusted}

We can now see the automatic change of dimensionality
of the field along the flow.
In the chart $U_1$, that is more appropriate to describe the 
low energy physics,
the field has dimension of mass and $Z_1$ is dimensionless.
Near GFP$_1$ the anomalous dimension is small and negative.
However, if we follow any RG trajectory towards the UV,
as discussed in Sect \ref{sec:U1},
the anomalous dimension graows and eventually tends to $-2$.
Recalling that the canonically normalized field 
$$
\tilde\phi=\sqrt{Z_1}\phi
$$
has scaling dimension $(d-2+\eta_1)/2$,
this means that the field is effectively dimensionless in the UV limit.
This is indeed the natural choice for the field
at GFP$_2$ in the chart $U_2$.
We observe that this automatic adjustment of the dimension
is a consequence of the form (\ref{betaZ2})
of the beta function of $\tilde Z_2$.
The fact that in the UV limit $\eta_1\approx -2$
also means that the wave function renormalization constant scales like 
$Z_1\sim k^2$ in that limit.
Thus in going from GFP$_1$ to GFP$_2$, $Z_1$ gets multiplied
by an infinite factor.

One can arrive at the same conclusions in the chart $U_2$
by observing that (\ref{betaZ1h}) can also be written in the form
\be
\beta_{\hat\zeta_1}=-(2+\eta_1)\hat\zeta_1\ ,
\ee
where 
\be
\eta_1=
\frac{8\hat\gamma(2+\hat\zeta_1)}{\hat\gamma+128\pi^2(1+\hat\zeta_1)^2}\frac{1}{Z_1}\ .
\ee


\subsubsection{The picture in dimensionful variables}

Even though the charts $U_1$ and $U_2$ are defined only for
the dimensionless coordinates on theory space,
when we consider the flow in the dimensionful parameters
appearing in the Lagrangian,
there is still a vestige of these coordinates
in the choice of the dimension of the field.

Because terms with fewer derivatives dominate at low energy,
it is natural to describe the IR physics in terms of the
dimensionful field $\phi$ with two-derivative kinetic term. 
In the limit $k\to 0$ both $Z_1$ and $g$ become constants,
see Equations (\ref{bZ1},\ref{bg}),
and recall that $Z_2$ is also a constant.
Therefore the effective action is (\ref{action1})
with arbitrary fixed coefficients.
This does not look like a free theory,
but if we identify the scale $k$ with a 
characteristic external momentum $p$,
by mere momentum counting
the first term is the dominant one in the IR limit.
The interaction is of order $g p^4$
and goes to zero much faster,
and the same happens for the higher-derivative kinetic term.
\footnote{The identification $k=p$ 
is unambiguous for the two point function,
but may require further qualifications for more complicated
physical processes.}
It is noteworthy that in order to identify the
$k\to 0$ limit as a free theory it is necessary to identify
$k$ as a physical momentum scale.

On the other hand, using the asymptotic behavior (\ref{asymp}),
and solving the flow equation for $\tilde Z_2$, we find that
the behavior for large $k$ is
$\tilde Z_2\sim \frac{11}{4}\log k$ and therefore
\footnote{This gives the anomalous dimension
$-2+1/\log{k}$.}
$$
Z_1= \frac{4Z_2 k^2}{11\log k}\ ,
$$
where $Z_2$ is fixed and arbitrary.
Then using (\ref{asymp}) one obtains
$$
g= -\frac{64\pi^2 Z_2^2}{11\log k}\ .
$$
Thus for $k\to\infty$, $g$ goes to zero
and we remain with a free theory.
Identifying again $k$ with the momentum in the two-point function,
the four-derivative kinetic term has an overall
momentum-dependence $p^4$ whereas the two-derivative one
goes like $p^4/\log{p}$.
Thus in the UV limit the 
four-derivative kinetic term is the dominant one,
but only logarithmically.

For large momentum it is natural to redefine the field 
as in (\ref{redef}), thus absorbing in the field the power in the running of $Z_1$ at high energy.
Then we find that $\hat\zeta_1=4/(11\log k)$ and
$$
\zeta_1=\frac{4\zeta_2 k^2}{11\log k}\ ,
$$
where $\zeta_2$ is an arbitrary dimensionless
constant that can be set to one.
Thus the ``mass squared'' $\zeta_1$ has the expected power behavior, with a logarithmic correction.
For the coupling $\gamma$ we get
$$
\gamma= -\frac{64\zeta_2 \pi^2}{11\log k}\ ,
$$
which is the expected behavior of a renormalizable coupling
and demonstrates asymptotic freedom at high energy.

If we now look at the IR limit using the field $\varphi$
we find that $\zeta_1$, $\zeta_2$ and $\gamma$
become all constants and we recover the previous statement that
the (free) two-derivative term is the dominant one.
Once again, the understanding of this limit
as a free theory hinges on identifying the RG scale $k$
with a characteristic external momentum $p$.

\subsubsection{The mass of the ghost}

In this section we think of the theory in Minkowski space,
where the classical action differs from (\ref{ct}) by an overall sign.
\footnote{We use signature $-+++$.}
It describes two propagating particles:
a normal massless scalar and a massive scalar ghost
with mass $m_{gh}^2=Z_1/Z_2$
(Notice that this statement is independent of the 
dimension of the field).
If $g=0$, $Z_1$ and $Z_2$ do not run and the statement
holds {\it verbatim} also at the quantum level.
If we now switch on $g$, the value of the physical (pole) mass
will receive quantum corrections.
The two-point function of the theory is defined as the
limit for $k\to 0$ of the two point function of the EAA.
In general, the dependence of the $n$-point functions on
the external momenta and on the parameter $k$ are not interchangeable,
but in the case of the two-point function,
given that the external momentum $p$ has the effect of an IR cutoff
in the integrations over the loop momenta,
the $k$-dependence is a good proxy for the $p$-dependence.
We can therefore reliably calculate the pole mass from
the running of the parameters $Z_1$ and $Z_2$
with the cutoff scale $k$.

In the presence of a running (renormalized) mass $m_R^2(k)$, 
the pole mass can be defined by
$$
m_{pole}^2=m_R^2(k=m_{pole})
$$
and corresponds to the threshold discussed in the previous section.
There is an old argument that if $m_R$
grows sufficiently fast, there may be no pole at all.
\footnote{See e.g. \cite{Floreanini:1994yp}.}
This would remove the unwanted ghost state.

Working in the chart $U_1$, 
the location of the pole is defined by 
$$
\tilde Z_2=1\ .
$$
Since all the trajectories run from $\tilde Z_2=\infty$
to $\tilde Z_2=0$, they inevitably hit the pole
and the argument mentioned above cannot apply.
However, the pole may be shifted to arbitrarily high scale.

To see this we start by setting $Z_1=1$ in the IR limit.
Then on the ``classical'' RG trajectory (\ref{ct}),
$Z_1=1$ everywhere and the pole mass is at
$$
k_P^2=\frac{Z_1}{Z_2}=\frac{1}{Z_2}\ .
$$
Let us now switch on the interaction.
The anomalous dimension $\eta_1$ is negative
and therefore $Z_1$ becomes larger than one.
Thus the pole is shifted to a larger value, 
compared to the ``classical'' trajectory.
This effect becomes stronger as one considers
trajectories that are further away from the classical one.
In the limit, consider a trajectory that is close to the separatrix.
Already for small $k$, $\tilde g$ becomes quickly very negative,
until one gets close to NGFP$_1$.
There the running of $\tilde g$ almost stops,
but $Z_1$ grows like
$$
Z_1(k)\sim k^{0.90}\ .
$$
This behavior can last for many orders of magnitude of $k$.
By the time the RG trajectory finally leaves the vicinity
of NGFP$_1$ and reaches $\tilde Z_2=1$, $Z_1$ can be
arbitrarily large.
Thus the mass of the ghost grows continuously
from $1/\sqrt{Z_2}$ to infinity as one moves 
from the classical RG trajectory to the separatrix.

\subsubsection{Redundant couplings and the essential RG}

One says that a coupling in the Lagrangian is ``redundant''
or ``inessential'' {\it at a specified FP},
if it can be removed from the Lagrangian by means
of a local field redefinition \cite{wegner,weinberg}.
There has been recently an interesting discussion of
the ``essential RG'', which is a way of simplifying the
RG flow by eliminating all redundant couplings \cite{Baldazzi:2021ydj}.

The prime example of a redundant coupling is the wave function
renormalization:
the parameter $Z_1$ is redundant at GFP$_1$ and
$\zeta_2$ is redundant at GFP$_2$,
because they can be fixed to $1$ by rescaling the field.
We have already taken this into account by putting suitable
powers of $Z_1$ or $\zeta_2$ in the definition
of the coordinates in theory space.
It is shown in Appendix A that doing so is necessary if we
demand that the beta functions be independent of
the dimension of the field.

However, also the parameter $Z_2$ is redundant at GFP$_1$.
Indeed, if $\tilde Z_2$ is infinitesimal,
it can be removed by an infinitesimal field redefinition
of the form 
$$
\delta\phi=\frac{Z_2}{2 Z_1}\Box\phi\ .
$$
One could therefore eliminate also $Z_2$ and get 
the essential flow equation for the single (in our approximation)
coupling $\tilde g$: in Figure 1, left panel,
it would be a flow in the vertical direction only,
and would lead to different properties of NGFP$_1$.
Similar considerations can be used to prove,
in a much more general setting than a mere scalar theory,
that if the kinetic term is the standard one
containing two derivatives, then in perturbation theory
one will never generate higher-derivative kinetic terms
\cite{Anselmi:2002ge}.

On the other hand when one considers GFP$_2$,
$\zeta_1$ is not redundant there because it cannot be removed by
a local field redefinition.
We conclude that by studying only
the essential RG at GFP$_1$ we would not realize
the possibility of flowing to GFP$_2$ in the UV,
which would imply an increase in the number of
propagating degrees of freedom,
but we would still have the possibility of flowing
to the non-Gaussian FP \cite{Baldazzi:2021ydj}.

\section{The flow from GFP$_1$ to GFP$_0$}

We now show that the preceding results are closely related
to a more familiar example.
Here we consider the case of a theory with an action of the form
\be
S=\int d^4x\left[
\frac12 Z_1(\partial\phi)^2
+\frac12 Z_0\phi^2
+\frac{1}{4!} \lambda \phi^4
+\ldots\right]
\label{action}
\ee
There are two potential terms that do not have shift symmetry.
The mass squared
has been called $Z_0$ since it is a member of the family
of quadratic Lagrangians.

\begin{figure}[h]
\begin{center}
\includegraphics[scale=0.45]{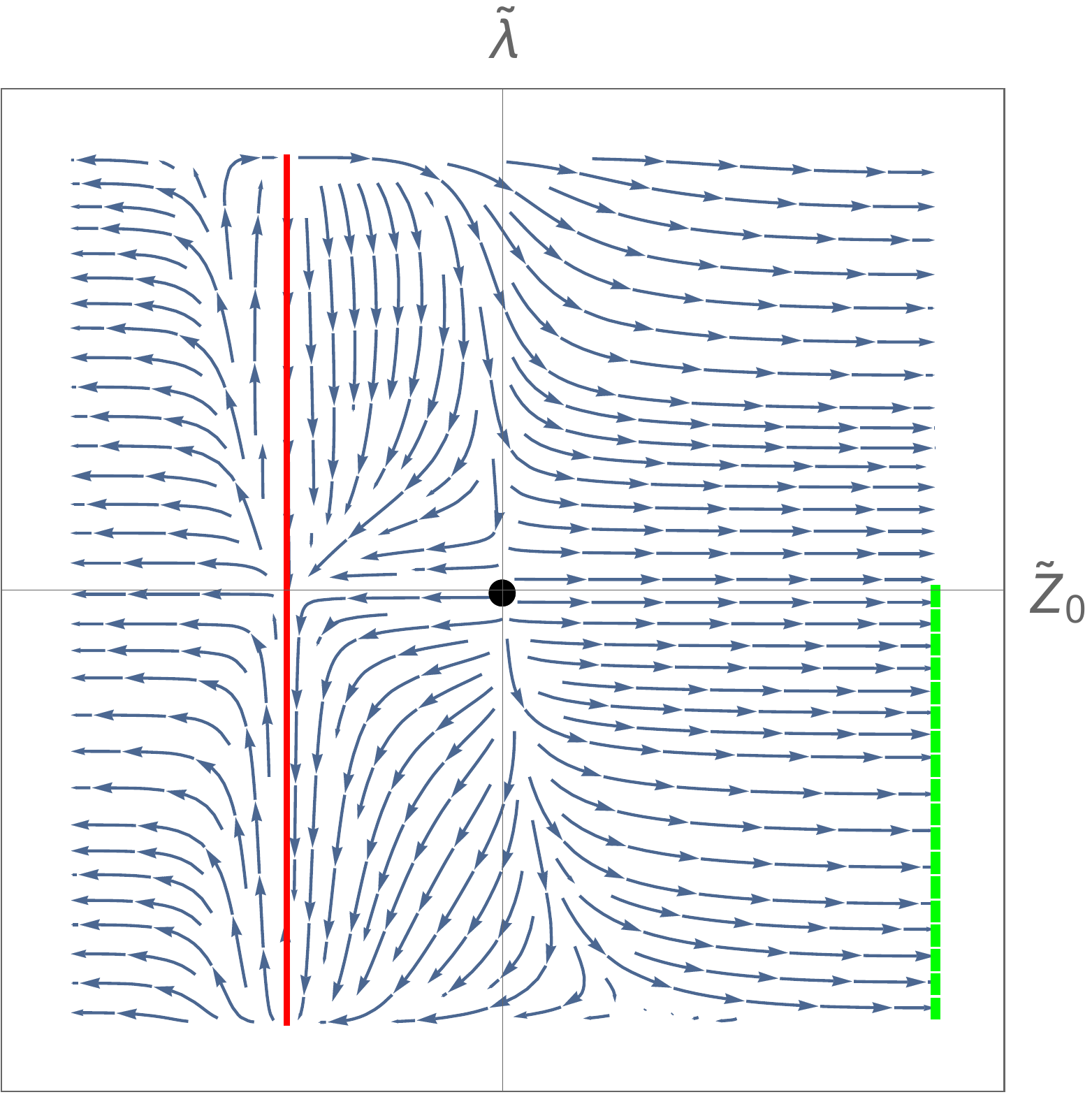}\quad
\includegraphics[scale=0.45]{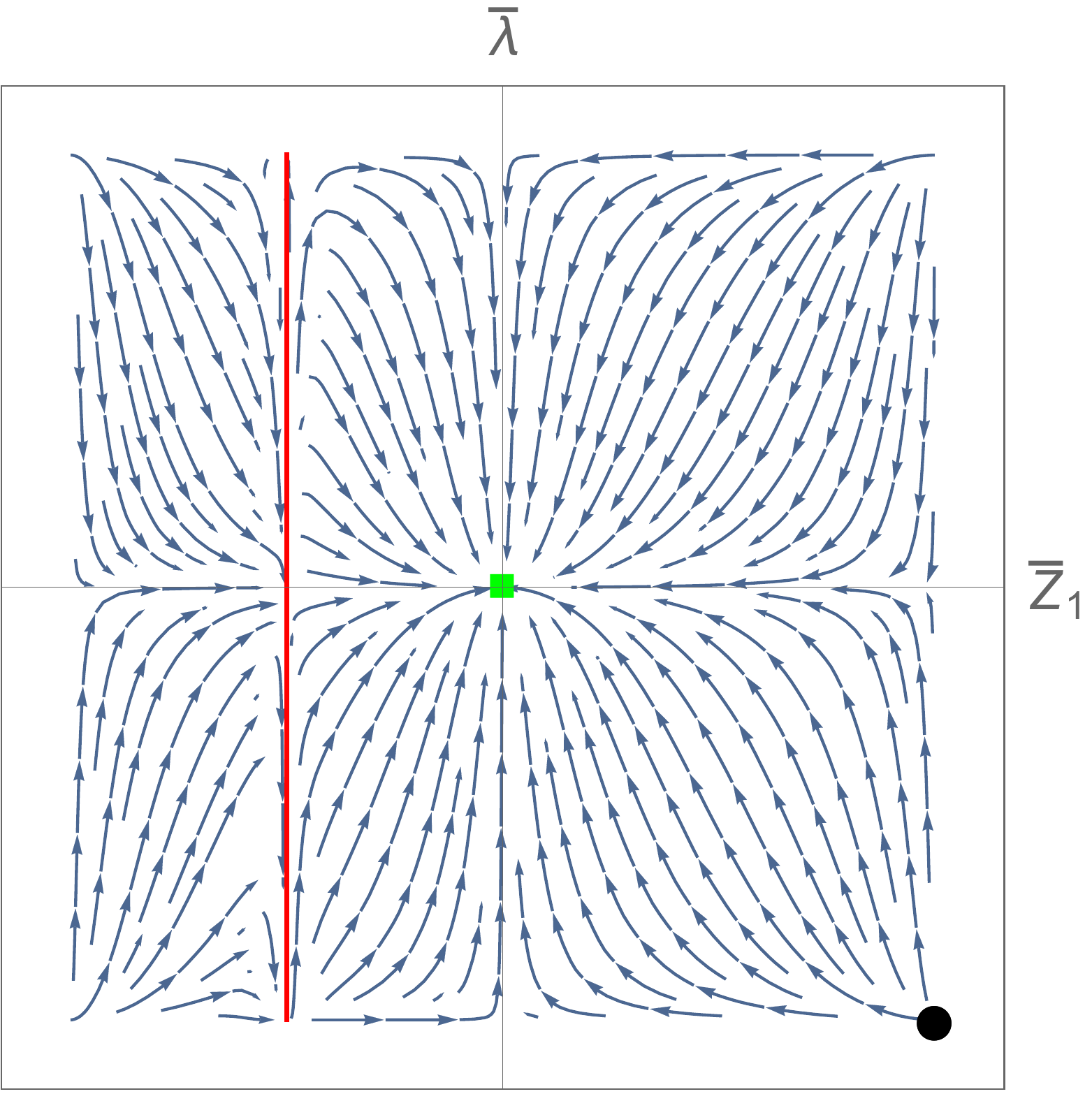}
\caption{The flow in the chart $U_1$ (left)
and in the chart $U_0$ (right).
Flow lines in the lower right quadrant go from GFP$_1$ (black)
to GFP$_0$ (green).
The vertical red lines are singularities of the flow.
}
\label{fig:pot}
\end{center}
\end{figure}

As in the case of the shift symmetric theories,
there are two natural choices of coordinates.
In the standard approach the field is assigned dimension of mass,
in which case the wave function renormalization $Z_1$
is redundant and the coordinates on theory space are
\be
\tilde Z_0=\frac{Z_0}{k^2 Z_1}
\ ,\qquad
\tilde\lambda=\frac{\lambda}{Z_1^2}\ .
\ee
This is the same chart $U_1$ considered above,
extended to a shift-non-invariant interaction,
and it has GFP$_1$ in the origin.
Using the cutoff $R_k(z)=Z_1(k^2-z)\theta(k^2-z)$,
the beta functions have the familiar form
\bea
\beta_{\tilde Z_0}&=&-2\tilde Z_0
-\frac{\tilde\lambda}{32\pi^2(1+\tilde Z_0)^2}\ ,
\\
\beta_{\tilde \lambda_0}&=&
\frac{3\tilde\lambda^2}{16\pi^2(1+\tilde Z_0)^3}\ .
\eea
The beta function of $Z_1$ is zero, and so is the anomalous dimension
$\eta_1=-\partial_t \log Z_1$.
There are no FPs in this chart except for GFP$_1$.
Using the rescaling $u=\tan(\tilde Z_0)$ and $v=\tan(\tilde\lambda)$,
the flow lines have the form shown in Figure \ref{fig:pot}.
We recognize that $\lambda$ is asymptotically free
for $\lambda<0$, as was noticed by Symanzik \cite{Symanzik:1973hx}. 
The fact that the flow lines asymptote horizontally is
due to the decoupling effect of the denominators:
for sufficiently small $k$, the running of $\lambda$ stops
whereas $\tilde Z_0$ continues to run due to the classical term.

Another chart $U_0$ is centered on the fixed point GFP$_0$
that is a free conformal field theory where the field $\phi$
has canonical dimension of mass squared.
This is sometimes called a trivial fixed point,
because in Minkowski signature it has no 
propagating degrees of freedom,
whereas in the Euclidean theory 
the correlation length at the fixed point is zero.
In this case the ``squared mass'' $Z_0$ is actually dimensionless
and redundant, $Z_1$ is an irrelevant coupling
of dimension $-2$ while $\lambda$ has dimension $-4$.
The coordinates on theory space are
\be
\bar Z_1=\frac{Z_1 k^2}{Z_0}
\ ,\qquad
\bar\lambda=\frac{\lambda k^4}{Z_0^2}\ .
\ee
The running of $Z_0$ is described by the anomalous dimension
\be
\eta_0=-\partial_t \log Z_0
=-\frac{\bar\lambda\bar Z_1}{32\pi^2(1+\bar Z_1)^2}
\ee
whereas the beta functions are
\bea
\partial_t \bar Z_1&=&(2+\eta_0)\bar Z_1\ ,
\\
\partial_t\bar \lambda&=&
(4+2\eta_0)\bar\lambda
+\frac{3\bar\lambda^2\bar Z_1}{16\pi^2(1+\bar Z_1)^3}\ .
\eea
Again, there are no nontrivial finite fixed points.
The beta function of $\bar Z_1$ can vanish either
because $\bar Z_1=0$, or because $\eta_0=-2$,
which is satisfied asymptotically for
$\bar\lambda\sim -64\pi^2\bar Z_1$ and $\bar Z_1\to\infty$.
These asymptotes correspond to GFP$_1$.

We note that, aside from the absence of other FP's,
the picture of the flow is very similar to the one
of the shift-symmetric theory.
In the chart $U_1$, the origin GFP$_1$
is the source of all flow lines with $\tilde\lambda<0$ and
GFP$_0$ corresponds to all points
with $\tilde\lambda<0$ finite and $\tilde Z_0\to\infty$,
so all the RG flow lines that are visible in the fourth quadrant
joint GFP$_1$ to GFP$_0$.
The same lines are visible in the chart $U_0$,
where GFP$_1$ is in the bottom right corner
and GFP$_0$ in the center.

The coordinate transformation between the charts $U_0$ and $U_1$ is
\be
\bar Z_1=\frac{1}{\tilde Z_0}\ ,\qquad
\bar\lambda=\frac{\tilde\lambda}{\tilde Z_0^2}\qquad
\mathrm{or\ conversely}\quad
\tilde\lambda=\frac{\bar\lambda}{\bar Z_1^2}\ .
\label{mappot}
\ee
and the beta functions transform as vectors under this transformation.

We observe that also in this case the kinetic term 
of the UV fixed point (GFP$_1$),
which gives rise to a propagating degree of freedom,
is a redundant operator from the point of view
of the IR fixed point (GFP$_0$),
where nothing propagates.
In fact, every local perturbation of GFP$_0$ is redundant.

\section{Discussion}

We now review our main findings and then comment
on possible extensions.

The general theory space contains all possible kinetic terms
and none of them plays an a priori preferred role.
It is only when one considers the perturbative expansion
around a Gaussian fixed point that the corresponding kinetic
term acquires a special meaning.
One then has a clear choice for the canonical dimensionality
of the field.
\footnote{In the case of ``generalized free theories'',
this is not the case.}
Otherwise, the choice of the dimension of the field
is essentially arbitrary:
physical conclusions are independent of this choice,
as we discuss in Appendix \ref{app:redef}.
However, the picture of the flow that follows from different
choices can be quite different.
For a fixed $k\geq 0$
the choice of kinetic term $Z_k\phi\Box^k\phi$ dictates
that the field has canonical dimension $(d-k)/2$
and this fixes the dimension of all the couplings $g_i$
in the Lagrangian.
When suitably rescaled by powers of $k$ and $Z_k$,
these couplings
define coordinates on an open subdomain $U_k$ of theory space.
Thus theory space is a manifold covered by infinitely many charts.
In the origin of the chart $U_k$ there sits GFP$_k$,
while all the other Gaussian FPs are outside this chart, 
but in its closure.

We have discussed mainly the RG trajectories joining GFP$_2$ to GFP$_1$.
They describe the unfamiliar situation of interacting theories 
that are free both in the UV and in the IR.
\footnote{For examples of gauge couplings in semisimple gauge theories 
that have this behavior see \cite{Bond:2017lnq}.}
Starting in the perturbative regime near GFP$_1$ at low energy, 
the coupling $\tilde g$ grows without bound 
as one goes towards the UV.
Superficially one may conclude that the theory does not have
a good UV limit.
However, one has to take into account the infinite amount of running 
of the wave function renormalization constants: 
while $\tilde g$ increases, $\tilde Z_2$ also increases
at a similar rate, in such a way that the combination
$\hat\gamma$ goes to zero, as seen from (\ref{mapinv}).
At the same time, the anomalous dimension also becomes
large and tends to $-2$, which is a sign that the
scaling dimension of the field becomes exactly zero,
as appropriate to GFP$_2$.
Thus, at some point, it becomes natural to move to the chart $U_2$
where one sees again a perturbative theory, this time governed
by the four-derivative kinetic term.

A very similar picture was found for the flow from GFP$_1$
to GFP$_0$, which corresponds to an asymptotically free
massive scalar theory {\it \`a la} Zymanzik,
that flows towards a trivial FP in the IR.
It is natural to conjecture the existence of flows between 
GFP$_k$ and GFP$_\ell$ with $k>\ell$.
However, theories with large $k$ have negative canonical
field dimension and infinitely many relevant couplings,
a problematic situation.
In spacetime dimension $d=4$ this already occurs for $k=3$,
and this is the reason why this case has not been discussed here.
One may think of restricting the number of relevant operators
by imposing higher order symmetries of the form
$$
\delta^{(k)}\phi=c^0+c^1_\mu x^\mu+\ldots
c^{k-1}_{\mu_1\ldots\mu_{k-1}}x^{\mu_1}\ldots x^{\mu_{k-1}}\ ,
$$
which is the symmetry of the kinetic term $\phi\Box^k\phi$,
for $k>0$.
More generally, this will be a symmetry for Lagrangians
where every field appears under at least $k$ derivatives.
By choosing the cutoff appropriately one can obtain flows
that respect the symmetry and therefore remain in
the symmetric subspaces of theory space,
which would acquire a complicated stratified structure.

All theories with $k>1$ have ghosts at perturbative level,
but we have shown that its mass depends on the trajectory
and there are trajectories where it is arbitrarily high.
The limiting case is the trajectory connecting GFP$_1$
in the IR to NGFP$_1$ in the UV, which is free of ghosts.
Another pathology 
is that the coupling must be negative,
leading to negative interaction energy.
This was already well known in the case of Symanzik's
asymptotically free scalar theory,
and it is generally agreed that in spite of the coupling
going asymptotically to zero, this is an unphysical feature
\cite{tHooft:1985mkt}.
Thus these theories are probably not very useful,
even as statistical models,
but we think that they still offer some interesting lessons
in quantum field theory.

Finally, let us discuss the limitations of our analysis.
We have kept all terms that are relevant at GFP$_2$,
so that the perturbative analysis is complete and 
self-consistent there:
the theories described by the action (\ref{action2}) 
lie on UV renormalizable (UV complete) trajectories,
because running the RG towards the UV they fall back onto GFP$_2$.
Any other deformation, when followed towards the UV,
will be pushed away from GFP$_2$ 
in some irrelevant direction and is not asymptotically free.
We have found that when we run the RG towards the IR,
all these trajectories tend to GFP$_1$.
However, midflows, we are in a strong coupling regime
and keeping only the terms in (\ref{action2})
(or equivalently (\ref{action1}))
becomes a drastic truncation.
Can we be sure that the conclusions cannot
be invalidated when we take into account additional interaction terms?
In fact, it is expected that when one departs from GFP$_2$,
all other couplings compatible with the symmetries will be
generated when one looks beyond linear order.
As an example one can consider the coupling $\gamma_2$
that multiplies the six-derivative operator 
$(\Box\phi)^2(\partial\phi)^2$.
The beta function of the dimensionless
$\hat\gamma_2=\gamma_2 k^2/ \zeta_2^2$ is
$$
\partial_t\hat\gamma_2=
2\hat\gamma_2+
\frac{1024\hat\gamma^2(2+\hat\zeta_1)}
{3(1+\hat\zeta_1)(\hat\gamma+128\pi^2(1+\hat\zeta_1)^2)}
+O(\hat\gamma_2^2)\ .
$$
As soon as $\hat\gamma$ is turned on, this beta function
becomes nonzero and $\hat\gamma_2$ starts to grow.
However, assuming that $\hat\gamma_2$ does not change too much 
the behavior of the other two couplings,
in the IR $\hat\zeta_1$ goes to infinity and
suppresses the loop term, while the classical term remains.
Thus $\hat\gamma_2$ is expected to go again to zero in the IR.
This is confirmed by numerically solving the equations.
Similarly, all the other local couplings will be generated,
but they are irrelevant for GFP$_2$ and even more so for GFP$_1$.
Thus, one expects that they all go to zero
as one flows towards the IR.

In the recent paper \cite{Laporte:2022ziz}, 
the shift-invariant scalar theory
has been studied in a truncation involving potentially infinitely
many terms, all powers of $(\partial\phi)^2$.
This gives more insight in the flow along the axis $\tilde Z_2=0$,
in particular on the properties of NGFP$_1$.
However, we observe that the term $(\Box\phi)^2$
will be generated by quantum fluctuations:
first, one loop effects of the coupling $\tilde g$
will generate the coupling $\gamma_2$ as indicated above
(this happens independently of the form of the kinetic
term and of the cutoff) and then one loop effects
involving $\gamma_2$ will give a nonzero beta function for $Z_2$.
\footnote{If one gives up $\mathbb{Z}_2$ symmetry,
$Z_2$ is generated by quantum fluctuations involving the
interaction $\partial_\mu\phi\partial_\nu\phi\partial^\mu\partial^\nu\phi$
\cite{steinwachs}.}
Since all the additional terms $((\partial\phi)^2)^n$, $n>2$,
are irrelevant at GFP$_2$, our conclusions 
will not be modified by the inclusion of such terms in the
truncation, except for changes in the properties of
the trajectories at strong coupling, and in particular
near the fixed point NGFP$_1$.

\vskip1cm

\noindent
{\bf Acknowledgements}
We thank G.P. Vacca, T. Morris, D. Litim and M. Reuter
for useful discussions at various stages of this work.

\appendix

\section{The dimension of the field is immaterial}\label{app:redef}

Assume that the effective action is a quasi-local functional
of the field of the form:
\be
\Gamma_k[\phi]=\sum_i g_i \cO_i(\phi)
\ee
For the sake of power counting,
the operators $\cO_i$ have the general form
\be
\cO_i(\phi)=\int d^dx\,\partial^{m_i}\phi^{n_i}\ ,
\ee
where the integrand stands for any scalar constructed
with $m_i$ derivatives and $n_i$ fields.
Assuming that the field has dimension $[\phi]=\df$,
$\cO_i$ has dimension $[\cO_i]=-d+m_i+n_i\df\equiv-d_i$
and the coupling $g_i$ has dimension $d_i$.

Now let us change variable from $\phi$ to
a new field $\phi'$ of dimension $\df'$:
\be
\phi'=\phi k^{\Delta\df}\ ,
\label{rescale}
\ee
with $\Delta\df=\df'-\df$.
The effective action of the new field is related to
that of the old field by
\be
\Gamma'_k[\phi']=\Gamma_k[\phi]\ .
\label{newgamma}
\ee
This means that while the two functionals have numerically
the same values when the fields are related as in (\ref{rescale}),
$\Gamma'_k$ is a different functional of its argument from $\Gamma_k$.
In particular, writing
\be
\Gamma'_k[\phi']=\sum_i g'_i \cO_i(\phi')\ ,
\ee
we find that
\be
g'_i=g_i k^{-n_i\Delta\df}\ .
\label{enrico}
\ee

At this point its is important to understand that
once the functional $\Gamma'_k$ has been defined by (\ref{newgamma}),
we are free to attribute all the $k$-dependence to the
couplings $g'_i$ and to think of the field $\phi'$
as being $k$-independent.
If we do so, we can use (\ref{enrico}) as a change of coordinates,
but not to calculate the $k$-dependence of $g'_i$.
Indeed, when we extract the beta functions from the
generating functionals $\partial_t\Gamma_k$
and $\partial_t\Gamma'_k$,
where $t=\log k$,
in both cases we will keep the field fixed.
In this way we arrive at the following relation between the beta functions:
\be
\partial_t g'_i=\partial_t g_i k^{-n_i\Delta\df}\ .
\label{ernesto}
\ee
The apparent contradiction with (\ref{enrico})
is due to our keeping $\phi'$ fixed.
Thus the missing contribution is
compensated by the fact that at the same time
we also ignore the $k$-term in (\ref{rescale}).
Equation (\ref{ernesto}) expresses the fact that
the calculation of the loop corrections
is the same, for any dimension of the field,
up to an overall factor of $k$ that accounts
for the different dimensions.

In the discussion of RG flows and fixed points 
we must use the dimensionless variables
$$
\tilde g_i=g_i k^{-d_i}\ ,\qquad
\tilde g'_i=g'_i k^{-d'_i}\ .
$$
However, using (\ref{ernesto}), one finds that the beta functions
of these dimensionless variables are different, namely
\be
\partial_t\tilde g'_i=\partial_t g'_i k^{-d'_i}
-d'_i\tilde g'_i
=\partial_t\tilde g_i-n_i\Delta\df\tilde g_i\ .
\ee
This does not happen if we properly take into account the
normalization of the field.
Among the couplings $g_i$ there is the wave function
renormalization constant $Z$ or $Z'$, that has dimension
$d_Z=d-2-2\df$ or $d_Z'=d-2-2\df'$ respectively.
Let us therefore define
$$
\tilde g_i=g_i Z^{-n_i/2}k^{-d_i+n_i d_Z/2}\ ,\qquad
\tilde g'_i=g'_i Z^{\prime-n_i/2}k^{-d'_i+n_i d_Z'/2}\ .
$$
Note that
$$
-d'_i+n_i d_Z'/2=
-d+m_i+n_i\frac{d-2}{2}
=-d_i+n_i d_Z/2\ ,
$$
and then, if we use (\ref{enrico}), $\tilde g'_i=\tilde g_i$.
Equation (\ref{ernesto}) implies that
$\partial_t Z'=\partial_t Z k^{-2\Delta \df}$
and therefore
\be
\partial_t\log{Z'}=\partial_t\log Z\ .
\ee
So, using the preceding formulae,
\bea
\partial_t\tilde g'_i&=&\partial_t g'_i Z^{\prime-n_i/2}k^{-d'_i+n_i d_Z'/2}
+\left(-d+m_i+n_i\frac{d-2-\partial_t\log{Z'}}{2}\right)\tilde g'_i
\nonumber\\
&=&\partial_t g_i Z^{-n_i/2}k^{-d_i+n_i d_Z/2}
+\left(-d+m_i+n_i\frac{d-2-\partial_t\log{Z}}{2}\right)\tilde g_i
=\partial_t\tilde g_i\ .
\eea
Thus the flows of the dimensionless {\it and} 
canonically normalized couplings is the same, 
independently of the dimension of the field.
This underlines the importance of including the
redundant wave function renormalization constants
in the definition of the coordinates on theory space.

\section{Beta functions of the dimensionful couplings}

As mentioned in the preceding Appendix,
the beta functions of the original, generally dimensionful,
parameters in the Lagrangian are related as in (\ref{enrico}).
This is because the calculation of the loop contributions
is the same, up to a redefinition of the dimensions.
We can see this explicitly in the case of the shift-symmetric theory.
Using a dimension-one field and action (\ref{action1}),
the beta functions are
\bea
\partial_t Z_1&=&
-\frac{(8-\eta_1)Z_1+16 k^2 Z_2}{128\pi^2(Z_1+k^2 Z_2)^2}\,g k^4
\label{bZ1}
\\
\partial_t Z_2&=&0
\\
\partial_t g&=&\frac{(10-\eta_1)Z_1-20 k^2 Z_2}{64\pi^2(Z_1+k^2 Z_2)^3}\,g^2 k^4
\label{bg}
\eea
With dimensionless field and action (\ref{action2}),
the beta functions are the same,
with the replacements $Z_i\to\zeta_i$, $g\to\gamma$.

\section{Other cutoffs}
We considered also cutoffs different from $R^{(24)}_k$ in (\ref{cut12}):
either the two-derivative cutoff:
\be
R_k^{(2)}=Z_1 (k^2-q^2)\theta(k^2-q^2)\ ,
\label{cut1}
\ee
which is the standard choice,
or the four-derivatives cutoff:
\be
R_k^{(4)}=Z_2 (k^4-q^4)(k^4-q^4)\ ,
\label{cut2}
\ee
which is sometimes used when the kinetic term
has four derivatives.

\subsection{Dimensionful fields and two-derivative cutoff}

With the field dimension set to 1 
and the cutoff $R^{(2)}_k$, we find
\be
\eta_1=
\frac{-6\tilde Z_2
+6\sqrt{\tilde Z_2}(1+\tilde Z_2)\arctan(\sqrt{\tilde Z_2})
}{
(1+\tilde Z_2)
\left(64\pi^2\tilde Z_2^2
+3\tilde g\sqrt{\tilde Z_2}\arctan(\sqrt{\tilde Z_2})
-3\tilde g\log(1+\tilde Z_2)\right)
}\,\tilde g
\ ,
\label{eta1sfull}
\ee
while the beta function is
\be
\beta_{\tilde g}=
(4+2\eta_1)\tilde g
+\left[\frac{5\left(4\tilde Z_2^2-(3+5\tilde Z_2+2\tilde Z_2^2)\eta_1
\right)}{128\pi^2\tilde Z_2^2(1+\tilde Z_2)^2}
+\frac{15\eta_1}{128\pi^2\tilde Z_2^{5/2}}\arctan\sqrt{\tilde Z_2}
\right]\tilde g^2
\ .
\label{betagetafull}
\ee
where (\ref{eta1sfull}) has to be used.
These are real only if $\tilde Z_2\geq0$, hence this set of flow equation has a restricted domain. 
Concerning the beta function of $Z_2$, we have $\partial_t Z_2=0$, hence the equation is independent of the cutoff and we ave again
\be
\beta_{\tilde Z_2}\equiv\partial_t\tilde Z_2
=(2+\eta_1)\tilde Z_2\ .
\label{betaZ2a}
\ee
In the limit $\tilde Z_2\to 0$, we recover the same nontrivial NGFP$_1$ and NGFP$_2$ of the cutoff $R^{(24)}_k$, while NGFP$_3$ is in the half-plane where the RG equations are complex.
In the right half-plane the condition $\eta_1=-2$ is satisfied on the curve 
\be
\tilde g=-\frac{64}{3}\pi \tilde Z_2^{3/2}.
\ee
The separatrix is not a straight line anymore, but we still have $\hat\gamma=0$ after the coordinate transformation from $U_1$ to $U_2$, hence a region with a flow between GFP$_1$ and GFP$_2$ exists also with this cutoff.

\subsection{Dimensionful field and four-derivative cutoff}

If we use the cutoff (\ref{cut2}),
the anomalous dimension is
\be
\eta_1=
\frac{3\tilde g\tilde Z_2\left(1+2\tilde Z_2\right)
}{8\pi^2(1+\tilde Z_2)}
-\frac{3\tilde g\tilde Z_2^2}{4\pi^2}\log\left(\frac{1+\tilde Z_2}{\tilde Z_2}\right)
\ .
\label{eta1sfull4}
\ee
and the beta function is
\be
\beta_{\tilde g}=
4\tilde g
+\frac{16+21\tilde Z_2(3+2\tilde Z_2)}{8\pi^2(1+\tilde Z_2)^2}\tilde g^2\tilde Z_2
-\frac{42}{8\pi^2}\tilde g^2\tilde Z_2^2
\log\left(\frac{1+\tilde Z_2}{\tilde Z_2}\right)
\ .
\label{betagetafull4}
\ee
where (\ref{eta1sfull4}) has been used.
The beta function of $Z_2$ is still given by (\ref{betaZ2a}).
The $\log\left(\frac{1+\tilde Z_2}{\tilde Z_2}\right)$ is ill-defined or complex in the region $-1<\tilde Z_2\leq0$, therefore the nontrivial FPs can not be found and
the only finite FP of these beta functions is GFP$_1$.

The beta functions are influenced by the logs on $\tilde Z_2=0$, but, up to a small neighborhood of this axis, the general behaviour of the RG flow in the fourth quadrant is very similar to the one described in section \ref{sec:U1}.
The condition $\eta_1=-2$ is satisfied for
\be
\tilde g\approx -16\pi^2\tilde Z_2\ .
\ee
giving exactly the same asymptotic behaviour as $R^{(24)}_k$.

\subsection{Dimensionless field and two-derivative cutoff}

The dimensionless field with the cutoff $R^{(2)}_k$ gives
\be
\beta_{\hat\zeta_1}=-2\hat\zeta_1
+\frac{6\hat\gamma\hat\zeta_1\left(
\sqrt{\hat\zeta_1}
-(1+\hat\zeta_1)\arccot\sqrt{\hat\zeta_1}
\right)}
{(1+\hat\zeta_1)\left(
64\pi^2\sqrt{\hat\zeta_1}
-3\hat\gamma\sqrt{\hat\zeta_1}\log\left(\frac{1+\hat\zeta_1}{\hat\zeta_1}\right)
+3\hat\gamma\arccot\sqrt{\hat\zeta_1}
\right)}
\ee
The beta function of $\hat\gamma$ is
{\small
\be
\beta_{\hat\gamma}
=\frac{5\hat\gamma^2\sqrt{\!\hat\zeta_1\!}
\left[
128\pi^2
+\hat\gamma\left(
6+9\hat\zeta_1
-6\sqrt{\!\hat\zeta_1\!}\,(4+3\hat\zeta_1)\arccot\!\sqrt{\!\hat\zeta_1\!}
+9\sqrt{\!\hat\zeta_1\!}\,(1+\hat\zeta_1)^2\!\arccot^2\!\sqrt{\!\hat\zeta_1\!}
-6\log\left(\!\frac{1+\hat\zeta_1}{\hat\zeta_1}\!\right)
\right)
\right]}
{64\pi^2(1+\hat\zeta_1)^2\left(
64\pi^2\sqrt{\hat\zeta_1}
-3\hat\gamma\sqrt{\hat\zeta_1}\log\left(\frac{1+\hat\zeta_1}{\hat\zeta_1}\right)
+3\hat\gamma\arccot\sqrt{\hat\zeta_1}
\right)}
\ee
}
while the anomalous dimension, which does not receive any contributions from one loop diagrams, is still zero, as in section \ref{sec:U2}.
Also with $[\varphi]=0$, the cutoff $R^{(2)}_k$ introduces some $\sqrt{\hat\zeta_1}$ that reduce the domain of reality of the beta functions to the half plane $\hat\zeta_1\geq0$. Moreover now there are also some logs which may give problems in the interval $-1<\hat\zeta_2\leq0$.
The qualitative behaviour in the bottom right quadrant is the same of the regulator $R^{(24)}_k$, with a separatrix leading to $FP_1$ at $\hat\zeta_1\to\infty$ and delimiting the attractive basin of GFP$_1$. The main difference is its trajectory near to GFP$_2$, which is not linear, as could be expected from the different asymptotic behaviour observed in $U_1$ for large $\hat Z_2$.

\subsection{Dimensionless field and four-derivative cutoff}

Using $R^{(4)}_k$, we have again $\eta_2=0$.
Then,
\be
\beta_{\hat\zeta_1}=-2\hat\zeta_1
+\frac{3\hat\gamma}{8\pi^2\hat\zeta_1^3}
\left[
-\frac{\hat\zeta_1(2+\hat\zeta_1)}{1+\hat\zeta_1}
+2\log(1+\hat\zeta_1)
\right]
\ee

\be
\beta_{\hat\gamma}
=\frac{5\hat\gamma^2}{8\pi^2\hat\zeta_1^4}
\left[
\frac{\hat\zeta_1(6+9\hat\zeta_1+2\hat\zeta_1^2)}{(1+\hat\zeta_1)^2}
-6\log(1+\hat\zeta_1)
\right]
\ee
The terms $\log(1+\hat\zeta_1)$ becomes complex for $\zeta_1<-1$, so these beta functions are real only in the region  $\zeta_1>-1$. The RG flow is similar to $R^{(24)}_k$ near GFP$_2$, but the behaviour for large $\hat\zeta_1$ is different: one can still observe a region where there is a flow from GFP$_2$ in the UV to GFP$_1$ at infinity in the IR, however there is no clear separatrix,
because with this regulator there is no NGFP$_1$ in the chart $U_1$.

\subsection{Vanishing regulator}

We may put a prefactor $a$ in front of the regulator (\ref{cut12}):
$$
R_k^{(24)}=a Z_1(k^2-q^2)\theta(k^2-q^2)
+a Z_2 (k^4-q^4)\theta(k^4-q^4)\ .
$$
The beta functions of dimensionful couplings in the action
depend on this parameter,
but the qualitative features of the flow are the same
for a large range of values of $a$.
For our discussion the main point to observe is that
the value of $\tilde g$ at NGFP$_1$ decreases when $a$ increases,
and increases when $a$ decreases.

The limit of vanishing regulator $a\to 0$
is interesting because it is related to dimensional regularization
\cite{Baldazzi:2020vxk,Baldazzi:2021guw}.
In this limit NGFP$_1$ goes to $-\infty$,
the point in the middle of the bottom side in Figure 1,
and the separatrix also disappears at infinity.
The trajectories we have been discussing now fill
up the bottom right quadrant
and the conclusions regarding the mass of the ghost remain valid.


\end{document}